\documentclass[manuscript]{aastex6}
\usepackage{graphicx}
\usepackage{subfigure}

\newcommand{\wind}{\textit{Wind}}
\newcommand{\stereoa}{\textit{STEREO A}}
\newcommand{\soho}{\textit{SOHO${/}$}LASCO}
\newcommand{\kms}{km s$^{-1}$}
\newcommand{\rsun}{$R_\sun$}
\usepackage{amsmath}
\shorttitle{Characteristics of 2018 August 20 CME }
\shortauthors{Chen et al.}

\begin{document}

\title{Characteristics of a Gradual Filament Eruption and Subsequent CME Propagation in Relation to a Strong Geomagnetic Storm}

\author{Chong Chen\altaffilmark{1,2}, Ying D. Liu\altaffilmark{1,2},
         Rui Wang\altaffilmark{1}, Xiaowei Zhao\altaffilmark{1,2}, Huidong Hu\altaffilmark{1},
     	 and Bei Zhu\altaffilmark{1}}  
         
\altaffiltext{1}{State Key Laboratory of Space Weather,
        National Space Science Center,
        Chinese Academy of Sciences, Beijing 100190, China;
        \href{mailto:liuxying@spaceweather.ac.cn}{liuxying@swl.ac.cn}}
\altaffiltext{2}{University of Chinese Academy of Sciences, Beijing 100049, China}

\begin{abstract}

An unexpected strong geomagnetic storm occurred on 2018 August 26, which was caused by a slow coronal mass ejection (CME) from a gradual eruption of a large quiet-region filament. We investigate the eruption and propagation characteristics of this CME in relation to the strong geomagnetic storm with remote sensing and in situ observations. Coronal magnetic fields around the filament are extrapolated and compared with EUV observations. We determine the propagation direction and tilt angle of the CME flux rope near the Sun using a graduated cylindrical shell (GCS) model and the Sun-to-Earth kinematics of the CME with wide-angle imaging observations from \stereoa. We reconstruct the flux-rope structure using a Grad--Shafranov technique based on the in situ measurements at the Earth and compare it with those from solar observations and the GCS results. Our conclusions are as follows: (1) the eruption of the filament was unusually slow and occurred in the regions with relatively low critical heights of the coronal field decay index; (2) the axis of the CME flux rope rotated in the corona as well as in interplanetary space, which tended to be aligned with the local heliospheric current sheet; (3) the CME was bracketed between slow and fast solar winds, which enhanced the magnetic field inside the CME at 1 AU; (4) the geomagnetic storm was caused by the enhanced magnetic field and a southward orientation of the flux rope at 1 AU from the rotation of the flux rope.

\end{abstract}


\section{Introduction} \label{sec:intro}

Filaments, which are also called prominences when they appear on the limb of the Sun, are cool and dense material suspended in the corona. They are usually located in highly sheared magnetic fields above polarity inversion lines (PILs). Filaments that occur over a wide range of latitudes outside active regions are called quiescent filaments. The filament field structure is probably a flux rope that supports the mass (e.g., \citealt{1989Priest}). A loss of stability or balance of the flux rope can cause an eruption, and the trigger mechanism for the eruption is a field of active research and still under debate. There are many trigger mechanisms, e.g., the catastrophe model \citep{1991Forbes}, torus instability \citep{2006Kliem}, kink instability \citep{2004Torok}, breakout model \citep{1999Antiochos}, shearing motion \citep{1990Aly} and flux emergence \citep{1995Feynman}. The decay index of coronal magnetic fields can be used to measure the torus instability. \citet{2006Kliem} find that the torus instability can be triggered if the value of the decay index n $\geq$ 1.5, while \citet{2010Demoulin} suggest that the critical decay index depends on the thickness and shape of the current channel. 

Filament eruptions can result in coronal mass ejections (CMEs; e.g., \citealt{2003Gopalswamy,1979Munro,2000Gilbert,2002Hori}). CMEs are large-scale magnetized plasma ejected from the Sun into interplanetary space and major drivers of space weather effects. CMEs and their upstream sheaths are responsible for most major geomagnetic storms in the terrestrial environment (e.g., \citealt{1988Tsurutani,1994Gonzalez, 2008Echer}). The properties and geoeffectiveness of CMEs have been widely studied and discussed. Previous studies of CME interplanetary propagation categorize CMEs into fast and slow events according to their speed in comparison with the average solar wind speed (e.g., \citealt{2000Gopalswamy})\footnote{Note that a slow CME with speed lower than the average solar wind speed (about 400 \kms) can also drive a shock when it is propagating into an even slower solar wind (e.g., \citealt{2016Liu, 2017Lugaz}). In this case, the speed of the ICME leading edge relative to the upstream solar wind speed has to be larger than the ambient fast magnetosonic speed (e.g., \citealt{2011Tsurutani}). Therefore, a shock ahead of an ICME observed at 1 AU is not necessarily suggestive of a fast CME near the Sun.}. During the transit from the Sun to Earth, fast CMEs experience an impulsive acceleration, and then rapidly decelerate to a nearly constant speed or gradual deceleration phase \citep{2013Liu}, while slow CMEs gradually accelerate out to about 20-30 \rsun{} and then travel with a nearly invariant speed around the average solar wind level \citep{2016Liu}. Fast CMEs have attracted great attention because of their association with interplanetary shocks, high energy particles, and intense geomagnetic storms upon arrival at the Earth. Slow CMEs are generally thought to be not as geoeffective as fast CMEs. However, slow CMEs take a long time to travel from the Sun to Earth, so they have a high potential to interact with other solar wind structures and their geoeffectiveness can thus be enhanced (e.g., \citealt{2016Liu, 2018He,2010Rouillard,2015Kataoka,2004Tsurutani}).

The interactions with the coronal magnetic fields and the solar wind structures can alter the propagation properties of CMEs. \citet{1986MacQueen}, \citet{2015Kay}, \citet{2015Wang}, and \citet{2018LiuYi} suggest that the propagation directions of CMEs can be influenced by the background coronal magnetic fields. Slow CMEs tend to deflect toward and propagate along streamer belts because of the strong open magnetic fields of coronal holes, while some fast CMEs may propagate away from streamer belts (e.g., \citealt{2009Xie, 2009Kilpua,2017Manchester} and references therein). The propagation directions of CMEs in interplanetary space can also be changed by interactions between CMEs or interactions with fast solar wind streams (e.g., \citealt{2001bGopalswamy, 2012Lugaz, 2014Liu, 2016Liu}), so it is difficult to predict the propagation direction in this case (e.g., \citealt{2009Echer}). In addition to deflections, CMEs could rotate during their journey from the Sun to Earth (e.g., \citealt{2007Yurchyshyn, 2010bLiu, 2010Lynch, 2011Vourlidas}). Although a statistical study finds that the tilt angles of CME flux ropes are close to those of the magnetic PILs in the corresponding solar source regions \citep{2015Marubashi}, cases with rotations larger than 100\degr{} are reported (e.g., \citealt{2005Rust, 2010bLiu, 2016Vemareddy}). The strength and duration of the southward magnetic field at 1 AU would change due to deflection and rotation of the CME flux rope, and as a result the geoeffectiveness of CMEs becomes uncertain. Thus, slow CMEs can also cause intense geomagnetic storms (e.g., \citealt{2004Tsurutani, 2018He}). However, the process of how slow CMEs leads to enhanced geoeffectiveness at 1 AU, compared with fast CMEs, is still unclear. 

On 2018 August 20, a large-scale quiet-region solar filament gradually erupted into a slow CME, which drove the third largest geomagnetic storm of solar cycle 24 with a $D_{st}$ minimum of $-$174 nT. This event provides a good opportunity to investigate how a slow CME can result in an intense geomagnetic storm. In this paper, we present a comprehensive analysis of this CME and its propagation from the Sun to Earth with remote sensing observations and in situ measurements. We describe the source region eruption in Section \ref{1}, the interplanetary propagation characteristics in Section \ref{2}, and the properties of the flux rope at 1 AU and the associated geoeffectiveness in Section \ref{3}. We conclude and discuss the results in Section \ref{4}.

\section{Source Region eruption} \label{1}

The filament was lunched around 08:00 UT on 2018 August 20, and weak flare ribbons and dimming regions were observed. Figure \ref{f1} shows an overview of the filament in a 193 \AA{} image from the Atmosphere Imaging Assembly (AIA; \citealt{2012Lemen}) with potential field source surface (PFSS) extrapolation results mapped on it. The filament with an ``L''·-like shape lies beneath a streamer with closed magnetic fields. The open magnetic field lines and the AIA image suggest that near the filament there are mainly two coronal holes, which can produce fast solar wind.

Figure \ref{f2}(a) shows an extreme ultraviolet (EUV) image of the filament at the time of the eruption. During the eruption, the top part of the filament began to oscillate and move upward. However, we can not trace the height of the filament radial motion. Flare ribbons became visible at about 08:00 UT on the bottom half of the filament and separated with a very slow velocity. However, the soft x-ray flux at 1--8 \AA{} from \textit{GOES} just increased sightly and did not reach class B. Dimming regions, which often appear in eruptions (e.g., \citealt{1997Sterling, 2007Sterling, 2005deToma}), were observed in Figure \ref{f2}(b). After the eruption, the northeast coronal hole gradually merged with a dimming region into a low-latitude coronal hole on 2018 August 21. This is probably the source of the fast solar wind following the interplanetary CME (ICME) in our case (see below). The filament channel is overlaid on the Helioseismic and Magnetic Imager (HMI; \citealt{2012Schou}) magnetogram in Figure \ref{f2}(c). The filament lies roughly along a neutral line with the positive polarity on the left side and the negative polarity on the right side, although it is not in an active region. The south end of the filament is located in the negative polarity region, which indicates that the axial magnetic field of the filament is likely pointing from northeast to southwest. The azimuthal magnetic field above the filament is from the left side to the right side. This magnetic field configuration indicates a left-handed flux rope around the filament.		

We give the distribution of the critical height at n $=$ 1.2 above the filament from the measurements of the decay index of coronal magnetic fields in Figure \ref{f3}. A decay index is defined as 
\begin{equation}
n=\frac{\partial{\ln B_h}}{\partial{\ln H}},
\end{equation}
where $B_h$ is the horizontal magnetic field component and $H$ is the height above the photosphere. \citet{2010Demoulin} show that for circular and straight current channels, which are deformable and as thick as those expected in the corona, the critical index is typically in the range [1.1,1.3]. Thus, we choose the value of 1.2 to calculate the distribution of the critical height. {Relatively low critical heights are seen in regions where the flare ribbons began. Therefore, we suggest that the bottom half of the filament erupted first due to the loss of stability and then perhaps destabilized the whole filament. 

We make a slice along the flare-ribbon separating direction (see Figure \ref{f2}) to create a distance-time diagram, which is shown in Figure \ref{f4}. From the distance-time plot, we find that the flare ribbons separated with a very slow speed of $\sim$3 \kms{} at the initial stage and $\leq$ 1 \kms{} later. The flare-ribbon separation lasted about 24 hours, which is a surprisingly long time period. These signatures indicate a gradual filament eruption, which will evolve into a slow and weak CME. The coronagraph observations from Solar and Heliospheric Observatory (\textit{SOHO}; \citealt{1995Domingo}) and Solar Terrestrial Relations Observatory (\textit{STEREO}; \citealt{2008Kaiser}) confirm a slow CME. Interestingly, the flare-ribbon separation overlaps with the slow acceleration of the CME (see below).

\section{Characteristics of Propagation} \label{2}

Coronagraph observations and corresponding GCS modeling are displayed in Figure \ref{f5}. The CME first appeared in the field of view of \stereoa/COR2 \citep{2008Howard} at 19:00 UT and \textit{SOHO}/C2 \citep{1995Domingo} at 21:00 UT on August 20. The CME signal looks very weak in \textit{SOHO} but much clearer in \stereoa{}, which indicates the importance to have a side view for CME observations. We use a graduated cylindrical shell (GCS) model proposed by \citet{2006Thernisien} to fit the CME based on running-difference coronagraph images from \stereoa/COR2 and \soho{}. The GCS model can determine the direction of propagation, tilt angle of CME flux rope and height (e.g., \citealt{2009Thernisien, 2010bLiu, 2017Hu, 2017Zhao}). Application of the model gives an average propagation direction of about 13\degr{} west of the Sun-Earth line and 6\degr{} north, which is consistent with the location of the flare ribbons (W12\degr{}N17\degr{}). The speed of the CME leading edge is accelerated from $\sim$70 \kms{} at 4.4 \rsun{} to $\sim$370 \kms{} at 16.2 \rsun{} (see below). According to the interval of the GCS CME fitting shown in Figure \ref{f4}, we find that the CME acceleration is coincident with the flare-ribbon separation. The tilt angle of the CME flux rope obtained from the GCS model is about 10\degr{} with respect to the ecliptic plane, which is quite different from the southwest orientation of the flare ribbons. We have actually tried to fit the CME flux rope with the same orientation as that of the flare ribbons, but a good match with \soho{} and \stereoa/COR2 coronagraph images can not be obtained simultaneously. Then, we choose a small tilt angle according to the loop-like structure in \stereoa/COR2 images and get a good visual consistency between the GCS model and observations.

Figure \ref{f6} shows the GONG synoptic magnetogram, over which the coronal magnetic field configuration and the heliospheric current sheet at 2.5 \rsun{} are plotted. The tilt angles of the white arrow and the lower cyan solid line suggest that the CME rotated in the low corona. We can not determine the direction and amount of the rotation, because the GCS model does not give the direction of the axial magnetic field of the CME flux rope. \citet{2007Green} find that for left-handed chirality a filament rotates counterclockwise. According to \citet{2007Green}, the CME in our study might have rotated counterclockwise by tens of degrees if the axis of the CME flux rope is westward, or by over 200\degr{} if the axis of the CME flux rope is eastward. However, we can not rule out the possibility that the CME has rotated clockwise by tens of degrees with the axis eastward. We compare the orientation of the heliospheric current sheet and the tilt angle of the CME flux rope by translating the lower cyan solid line to the location of the heliospheric current sheet along the same longitude. As shown in Figure \ref{f6}, the axis of the CME flux rope matches the tilt of the heliospheric current sheet, which supports the speculation of \citet{2008Yurchyshyn} that the axis of the ejecta may rotate in such a way that it locally aligns itself with the heliospheric current sheet.

We investigate the CME evolution in interplanetary space by producing a time-elongation map (e.g., \citealt{2008Sheeley, 2009Davies, 2010aLiu}). The time-elongation map shown in the left panel of Figure \ref{f7} is produced by stacking running-difference images of COR2, HI1, and HI2 from \stereoa{} within a slit along the ecliptic plane. The track associated with the CME can be identified from the map. Elongation angles of the CME leading edge in the ecliptic plane are extracted along the track, and then can be converted to radial distances using the methods summarized in \citet{2010bLiu}. Given that \stereoa{} is 108\degr{} east of the Sun-Earth line, we derive the CME kinematics using a harmonic mean (HM) approximation \citep{2009Lugaz}, which assumes that CMEs are attached to the Sun as a spherical front and move along a fixed radial direction. Readers are directed to \citet{2013Liu} for discussions of selection of CME geometry depending on the observation angle of the spacecraft.

The CME kinematics obtained from the GCS modeling and HM approximation are presented in the right panel of Figure \ref{f7}. Note that we have used the GCS propagation longitude (W13\degr{}) as input for the HM approximation. The GCS model and the HM approximation give consistent CME height and speed profiles below $\sim$17 \rsun. The CME was gradually accelerated to about 370 \kms{} below $\sim$17 \rsun{} and then moved with a nearly constant speed. This is a typical speed profile for slow CMEs, i.e., a gradual acceleration followed by a nearly constant speed around the average solar wind level \citep{2016Liu}. We use a linear extrapolation of the distances after 17 \rsun{} to predict the CME arrival time at the Earth, which is about 15:21 UT on August 25. We will compare the predicted arrival time and the CME final speed with in situ measurements at the Earth. Given the flux rope orientation and the slow speed, the CME would have little geoeffectiveness at the Earth.

\section{Properties at 1 AU} \label{3}

Figure \ref{f8} shows the in situ measurements from \wind{} associated with 2018 August 20 CME without a preceding shock. The flux rope interval is from 14:10 UT on August 25 to 09:09 UT on August 26, which is determined through a boundary-sensitive reconstruction technique (see below). It has all the three signatures of a magnetic cloud (MC), i.e., a smooth and strong magnetic field, a coherent rotation of the field, and a low proton temperature, according to the definition of \citeauthor{1981Burlaga} (\citeyear{1981Burlaga}; also see \citealt{2008Echer, 2019MengXing} and references therein). The CME arrival time (15:21 UT on August 25) at the Earth predicted by wide-angle imaging observations is in good agreement with the observed MC leading boundary time (14:10 UT). The average speed across the MC leading boundary is about 400 \kms, which is also well consistent with the predicted speed ($\sim$370 \kms{}). According to the velocity measurements, the CME is bracketed between slow and fast solar winds. The source of the fast solar wind is probably the coronal hole east of the filament, as discussed in section \ref{1}. Inside the MC, the magnetic field strength is as high as 19.1 nT, and the peak southward component reaches $-$16.4 nT. The magnetic field inside the CME must have been enhanced because the CME is inside a compression region between slow and fast solar winds. A similar situation is observed by \citet{2018He} and \citet{1989Tang}. The N component remains largely negative inside the MC, which indicates a southward orientation of the CME flux rope. This is different from the tilt angle derived from the GCS model, so the flux rope may have rotated again in interplanetary space. The $D_{st}$ index reaches a minimum of $-$174 nT, which is the third largest one of solar cycle 24. The largest geomagnetic storms of solar cycle 24 so far occurred on 2015 March 17 and June 22 with $D_{st}$ minima of $-$223 and $-$195 nT, respectively \citep{2015Liu}.

We reconstruct the flux rope at 1 AU using a Grad-Shafranov (GS) technique \citep{1999Hau, 2002Hu}, which has been validated by multi-spacecraft measurements \citep{2008Liu, 2009Mostl}. As shown in Figure \ref{f9}, the reconstruction results give a left-handed flux rope, which is consistent with the solar source observations. The elevation angle of the flux rope is about $-$51\degr{} and the azimuthal angle is about 299\degr{}. The flux-rope tilt angle is very different from that determined from the GCS model near the Sun, which indicates that the CME rotated again in interplanetary space. The southeast axis orientation of the flux rope explains the prolonged southward magnetic fields inside the MC, which is the most important trigger of geomagnetic storms. From the in situ measurements and the GS reconstruction results, the unexpected intense geomagnetic storm is mainly caused by the enhanced magnetic field in a solar wind compression region and a southward orientation of the flux rope at 1 AU from the rotation of the flux rope.

Figure \ref{f10} shows the heliospheric current sheet map at 1 AU at 14:10 UT on 2018 August 25 (adapted from \url{http://www.predsci.com/mhdweb/summary_plots.php}). The propagation direction of the CME is 16\degr{} west of the Sun-Earth line derived from the GCS model, and the Carrington longitude of the Sun-Earth line is 103\degr{} at 14:10 UT on 2018 August 25. Assuming that the propagation direction of the CME does not change much beyond 17 \rsun{}, the CME would arrive at 1 AU with a Carrington longitude of about 120\degr{} as marked by the diamond symbol in Figure \ref{f10}. It is clear that the CME rotated in interplanetary space. Again, the flux-rope rotation is such that the axis of the flux rope is aligned with the tilt of the local heliospheric current sheet at 1 AU.

\section{summary and discussions}\label{4}
   
We have investigated the characteristics of the 2018 August 20 CME in relation to an intense geomagnetic storm, using remote sensing observations from \textit{SDO}, \textit{STEREO}, and \textit{SOHO} and in situ measurements at \wind. The PFSS model together with EUV observations is used to examine the coronal magnetic field configuration and filament eruption. The Sun-to-Earth propagation of the CME is analyzed using the GCS model and wide-angle imaging observations. Finally, a GS reconstruction method is employed to understand the flux-rope structure and how the structure controls geomagnetic activity. The results are summarized and discussed below, which provide insights on how a weak CME can result in an unexpected strong geomagnetic storm. 

The filament eruption is a very slow process and lasted about 24 hours, as can be seen from the gradual separation of the flare ribbons. This indicates a slow CME, whose acceleration is coincident with the flare-ribbon separation. We give the distribution of the critical height above the filament from the measurements of the decay index of coronal magnetic fields. Regions with relatively low critical heights above the filament are found and are consistent with where the flare ribbons began. Therefore, we suggest that the part of the filament with relatively low critical heights first erupted due to the loss of stability and then destabilized the whole filament.

The axis of the CME flux rope rotated in the corona as well as in interplanetary space, which tended to be aligned with the local heliospheric current sheet. The solar source observations suggest that the orientation of the flux rope is southwest, but the tilt angle of the CME flux rope determined by the GCS model is largely horizontal. Therefore, the CME may have rotated in the low corona. The axis of the CME flux rope matches the tilt of the heliospheric current sheet at 2.5 \rsun{}. Through the reconstruction of the flux-rope structure near the Earth, we find that the elevation angle of the flux rope is about $-$51\degr{}, which is different from the GCS orientation. Therefore, the CME may have rotated again in interplanetary space, and the rotation is such that the tilt angle of the CME flux rope is aligned with the heliospheric current sheet at 1 AU. These results support the speculation of \citet{2008Yurchyshyn} that the axis of the CME flux rope may rotate in such a way that it aligns itself with the local heliospheric current sheet.

According to the in situ measurements, the CME was bracketed between slow and fast winds, which enhanced the magnetic field inside the CME at 1 AU. The magnetic strength inside the MC is as high as 19.1 nT, and the southward field component peaks at $-$16.4 nT. The fast solar wind following the CME from behind came from a coronal hole east of the filament. According to the GS reconstruction results, the prolonged southward magnetic field
inside the MC is mainly from the axial component of the largely southward inclined flux rope (e.g., \citealt{2016Hu}). If the flux rope had not rotated in interplanetary space, the strength and duration of the southward magnetic field would be much smaller. These results indicate that the unexpected intense geomagnetic storm was caused by the enhanced magnetic field of the CME in the solar wind compression region and a southward orientation of the flux rope at 1 AU from the rotation of the flux rope.

In summary, our analysis gives a whole process of the CME propagation from the Sun to Earth in connection with geomagnetic activity. The large-scale quiescent filament gradually erupted beneath a streamer, which evolved into a weak CME that entered a solar wind compression region between slow and fast winds. Meanwhile, the CME flux rope rotated in the corona as well as in interplanetary space and tended to be aligned with the local heliospheric current sheet. As a result, the CME arrived at 1 AU with the enhanced and prolonged southward magnetic field, which caused the third largest geomagnetic storm of solar cycle 24. These results indicate the crucial importance of understanding the physical process of CME evolution for accurate space weather forecasting.

\acknowledgments

The research was supported by NSFC under grants 41774179 and 41604146, and the Specialized Research Fund for State Key Laboratories of China. We acknowledge the use of data from {\textit{STEREO}, \textit{SDO}, \textit{SOHO}, \wind, and \textit{GONG}}. The MHD simulation results are obtained from \url{http://www.predsci.com/mhdweb/summary_plots.php}.

\clearpage
\begin{figure}
\epsscale{0.8}
\plotone{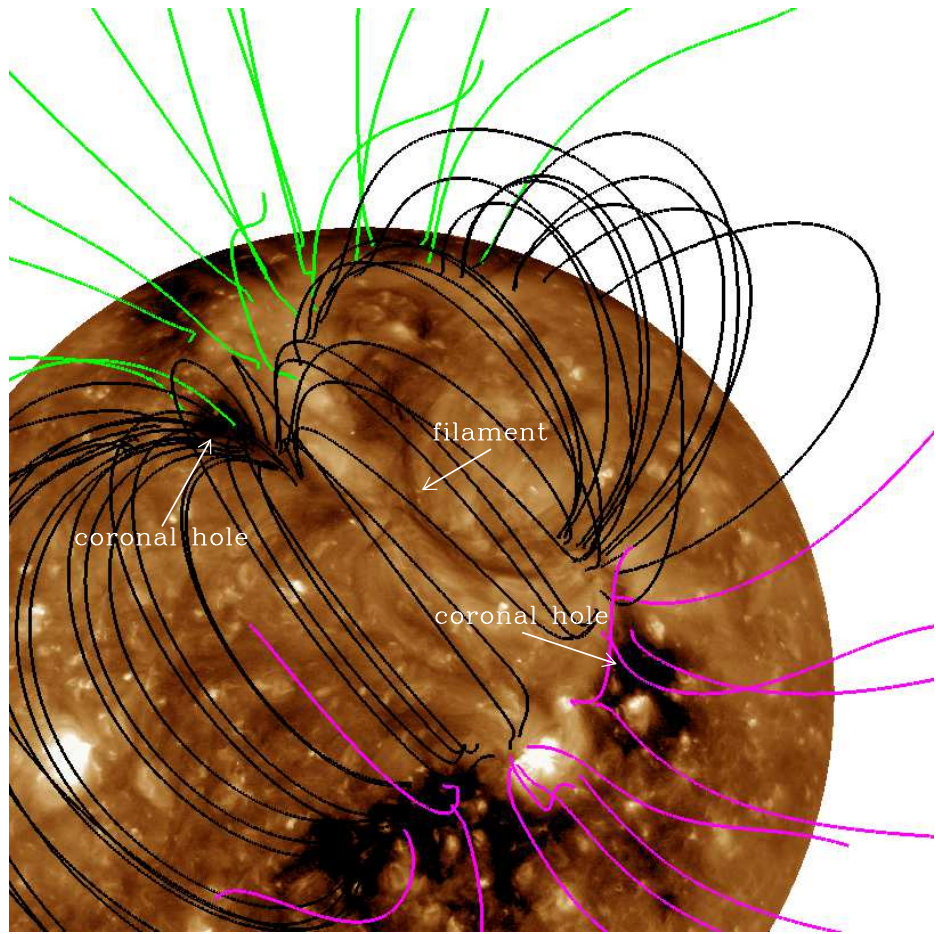}
\caption{\label{f1}PFSS modeled coronal magnetic fields
surrounding the source region mapped onto the AIA 193 \AA{} image at 04:00 UT on 2018 August 20. Closed field lines are in black, open positive field lines in green, and open negative field lines in magenta.} 
\end{figure}

\clearpage
\begin{figure}
\epsscale{1.2}
\plotone{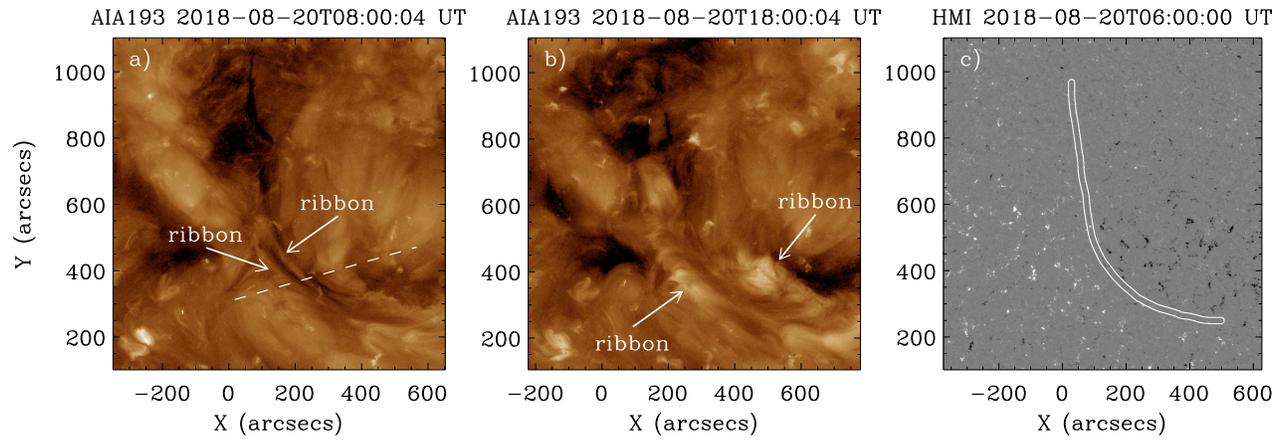}
\caption{\label{f2}EUV images at 193 \AA{} and magnetogram around the filament region. (a-b) EUV images at 08:00 and 18:00 UT on August 20 showing the filament eruption. The white arrows mark the flare ribbons. The white dash line indicates a slice along the flare-ribbon separating direction to create a distance-time diagram. (c) HMI magnetogram with mapped filament channel.}
\end{figure}

\clearpage
\begin{figure}
\epsscale{0.8}
\plotone{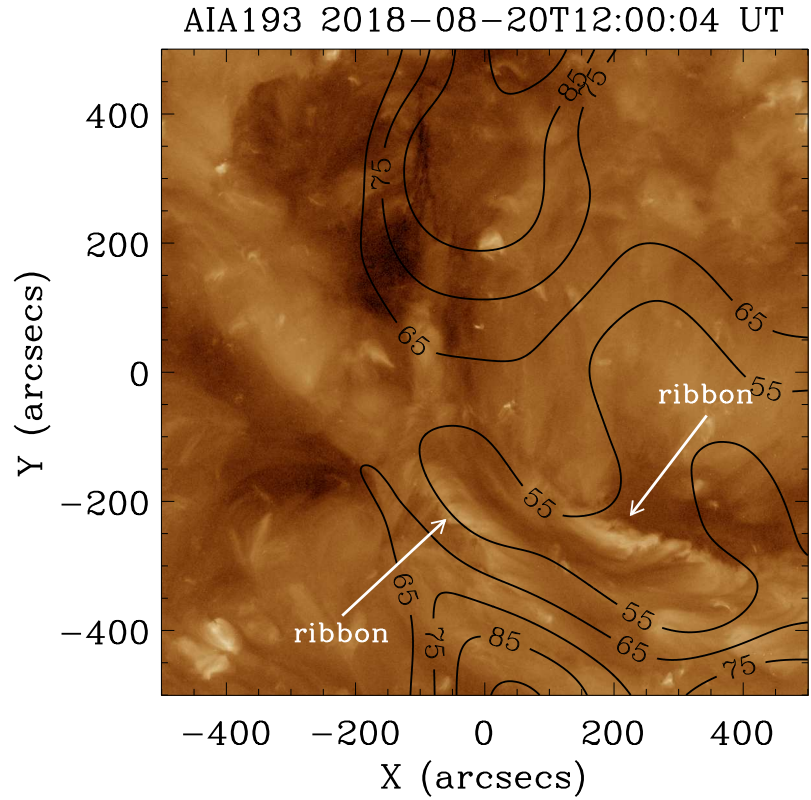}
\caption{\label{f3}Distribution of the critical height at n $=$ 1.2 on the AIA 193 \AA{} image at 12:00 UT on August 20 with contours of [55, 65, 75, 85] Mm. The white arrows indicate the locations of the flare ribbons.}
\end{figure}

\clearpage
\begin{figure}
\epsscale{1.1}
\plotone{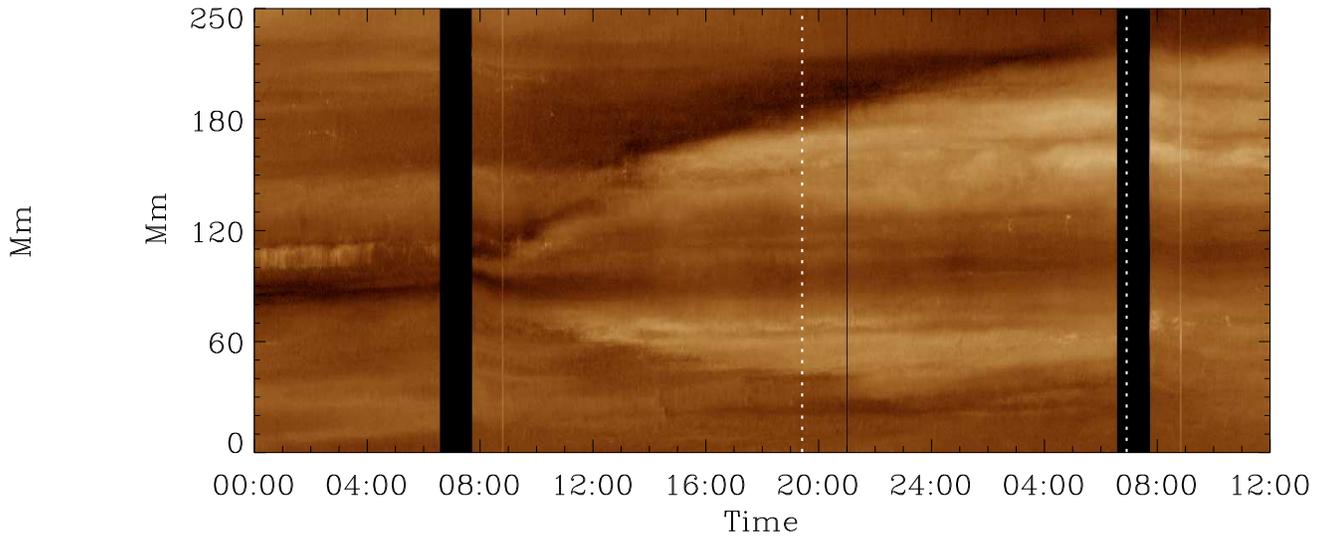}
\caption{\label{f4}Distance-time diagram of the flare ribbons created along the slice in Figure \ref{f2}(a). The white vertical dotted lines denote the interval of the GCS CME fitting.}
\end{figure}

\clearpage
\begin{figure}
\epsscale{0.8}
\plotone{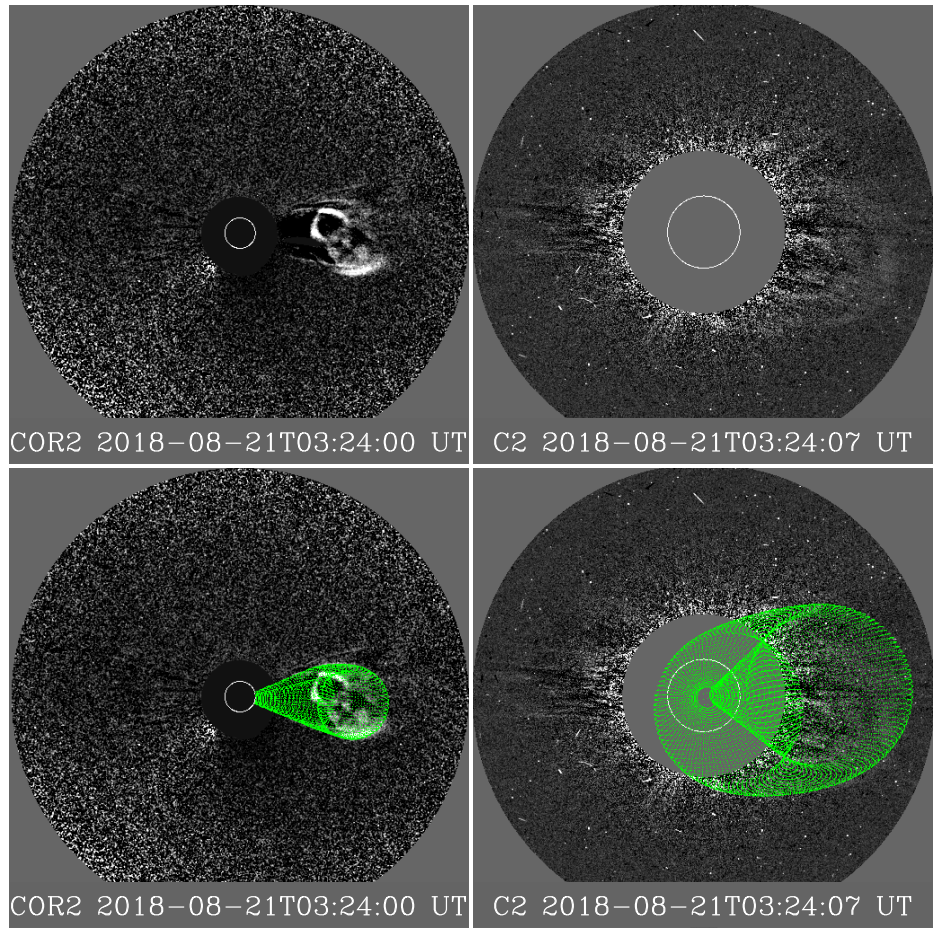}
\caption{\label{f5}Running-difference coronagraph images and corresponding GCS modeling (green grids) from \textit{LASCO} C2 (right) and \stereoa/COR2 (left). Detectors and times are given in the images. Note that \stereoa{} is 108\degr{} east of the Sun-Earth line.}
\end{figure}

\clearpage
\begin{figure}
\epsscale{1.2}
\plotone{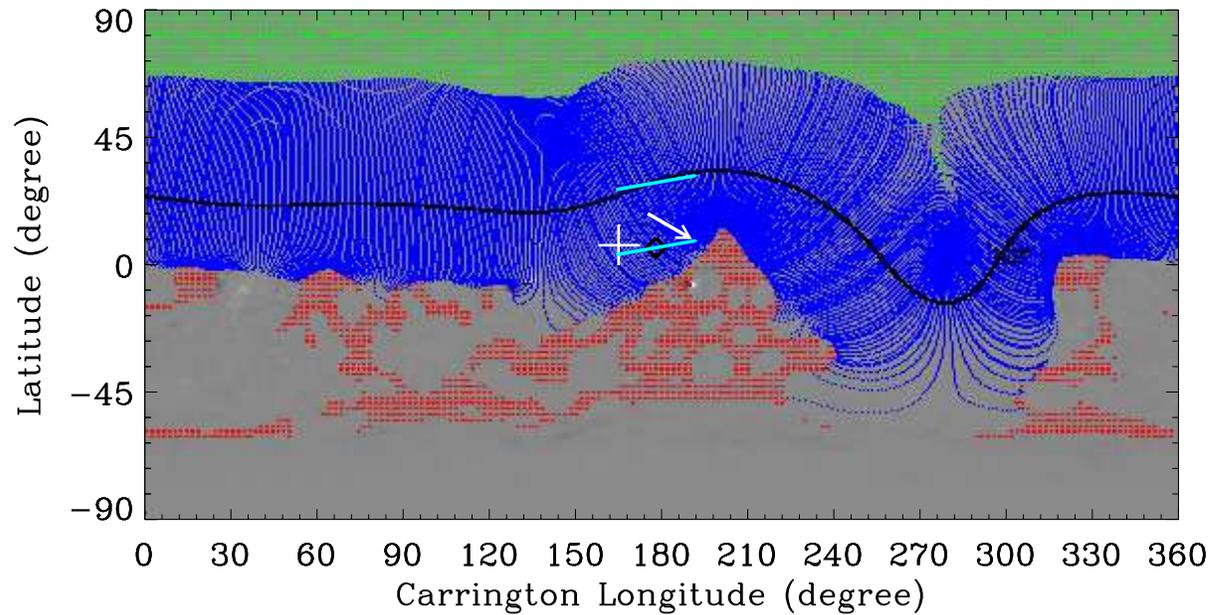}
\caption{\label{f6}GONG synoptic map for CR2207 with the PFSS magnetic field lines. Closed field lines are in blue, open positive field lines in green, and open negative field lines in red. The black solid line is the heliospheric current sheet. The white cross indicates the location of the Earth in Carrington coordinates at the time of 18:30 UT on 2018 August 20 (the start time of the GCS CME fitting). The orientations of the flare ribbons and the CME flux rope are marked by the white arrow and the lower cyan solid line, respectively. The diamond gives the propagation direction of the CME. The upper cyan solid line is a translation from the lower cyan solid line to the heliospheric current sheet along the same longitude.}
\end{figure}

\clearpage
\begin{figure}
\epsscale{2}
\centering
\subfigure{\includegraphics[width=3.4in]{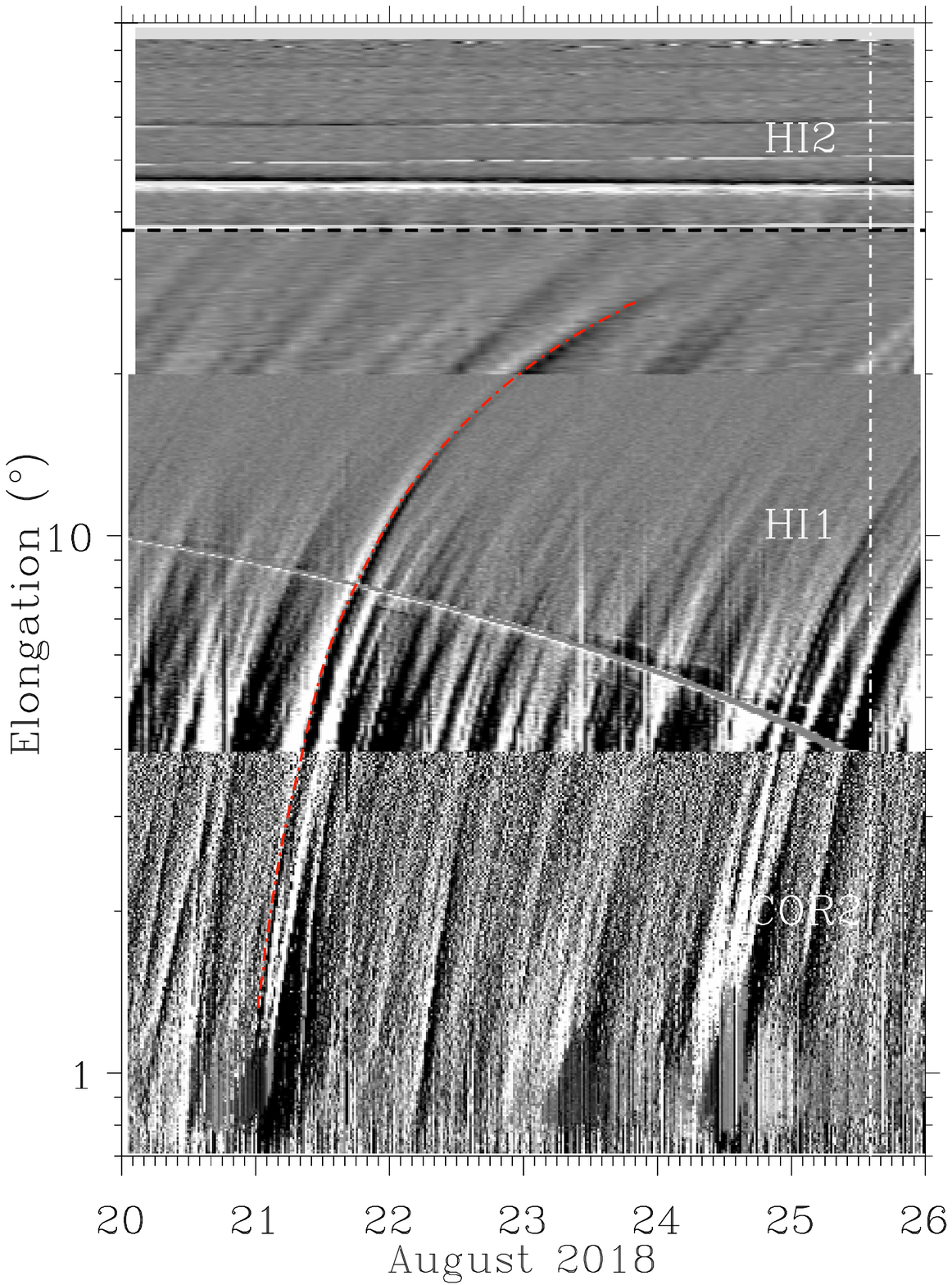}} 
\subfigure{\includegraphics[width=3.4in]{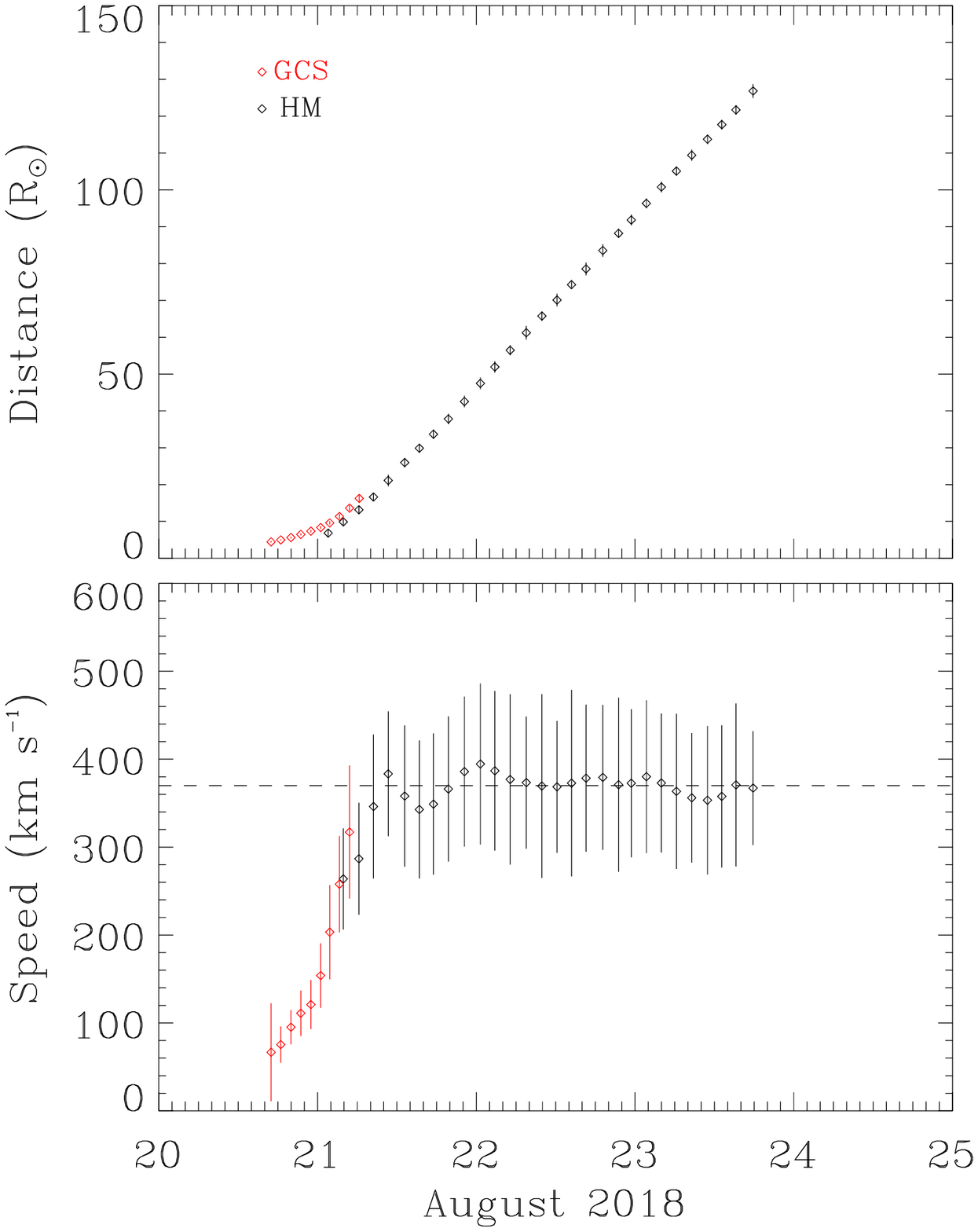}}
\caption{\label{f7}Left: time--elongation map constructed from running-difference images of COR2, HI1 and HI2 from \stereoa{} along the ecliptic plane. The red curve indicates the track of the CME, along which the elongation angles are extracted. The black horizontal line denotes the elongation angle of the Earth. The white vertical line shows the arrival time of the CME leading edge at the Earth. Right: radial distance and speed profiles of the CME leading edge derived from the GCS model (red) and the HM approximation (black). The horizontal dash line marks the average speed of $\sim$370 \kms{} beyond 17 \rsun{}.}
\end{figure}

\clearpage
\begin{figure}
\epsscale{0.8}
\plotone{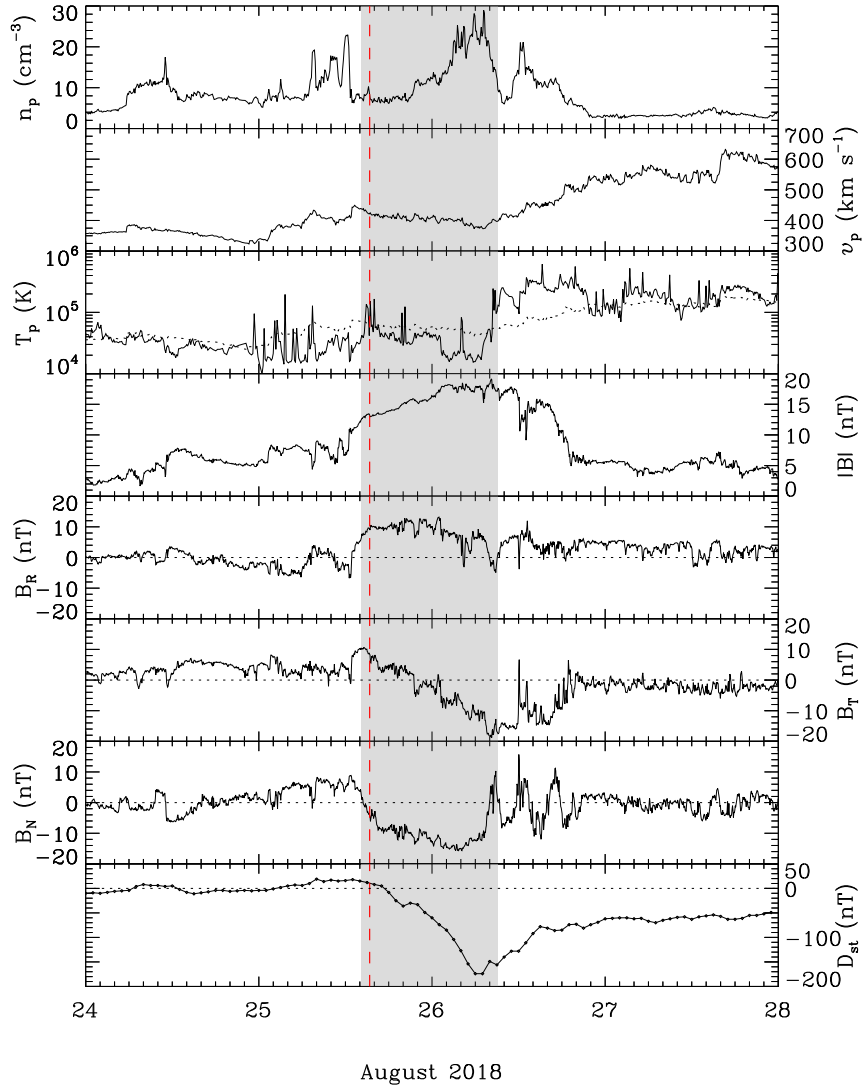}
\caption{\label{f8}Solar wind plasma and magnetic field parameters from \wind{} associated with the 2018 August 20 CME. From top to bottom, the panels show the proton density, bulk speed, proton temperature, magnetic field strength and components, and $D_{st}$ index, respectively. The dotted line in the third panel denotes the expected proton temperature calculated from the observed speed \citep{1987Lopez}. The shaded region indicates the flux rope interval determined by our GS reconstruction. The red vertical line represents the predicted arrival time of the CME leading edge at the Earth.}
\end{figure}

\clearpage
\begin{figure}
\epsscale{1.1}
\plotone{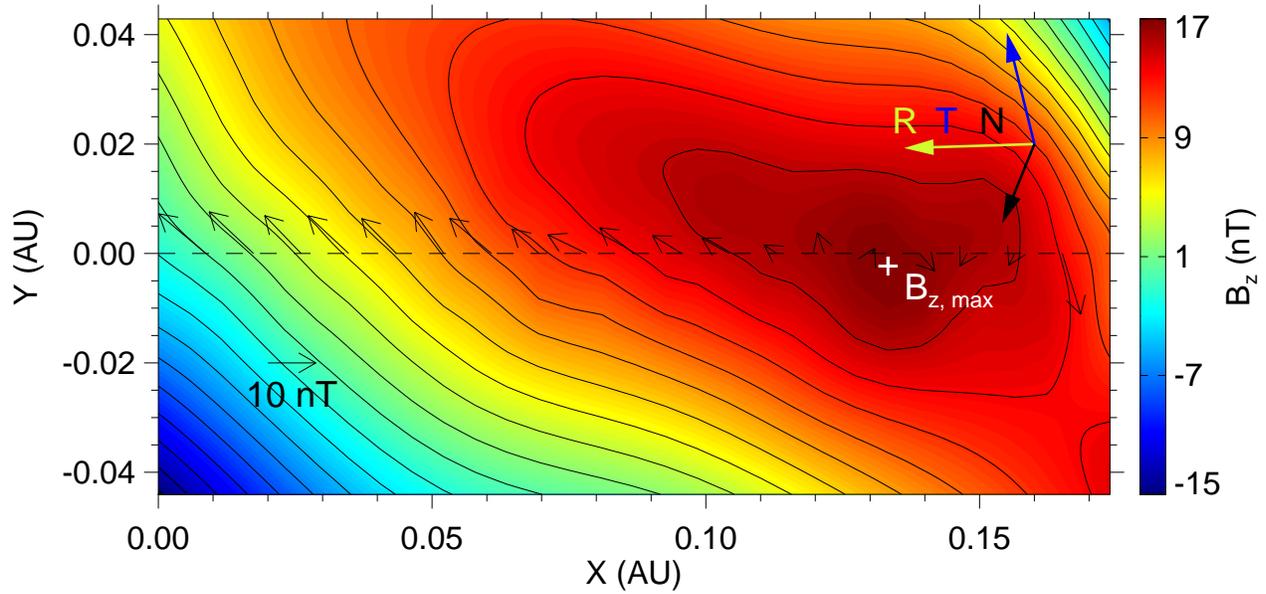}
\caption{\label{f9}Reconstructed cross section of the flux rope at \wind. Black contours are the distribution of the vector potential, and the color shading shows the value of the axial magnetic field strength. The location of the maximum axial field is indicated by the white cross. The horizontal dash line marks the trajectory of the \wind{} spacecraft. The thin black arrows denote the direction and magnitude of the observed magnetic field projected onto the cross section, and the thick colored arrows show the projected \textbf{RTN} directions.}
\end{figure}

\clearpage
\begin{figure}
\epsscale{1.1}
\plotone{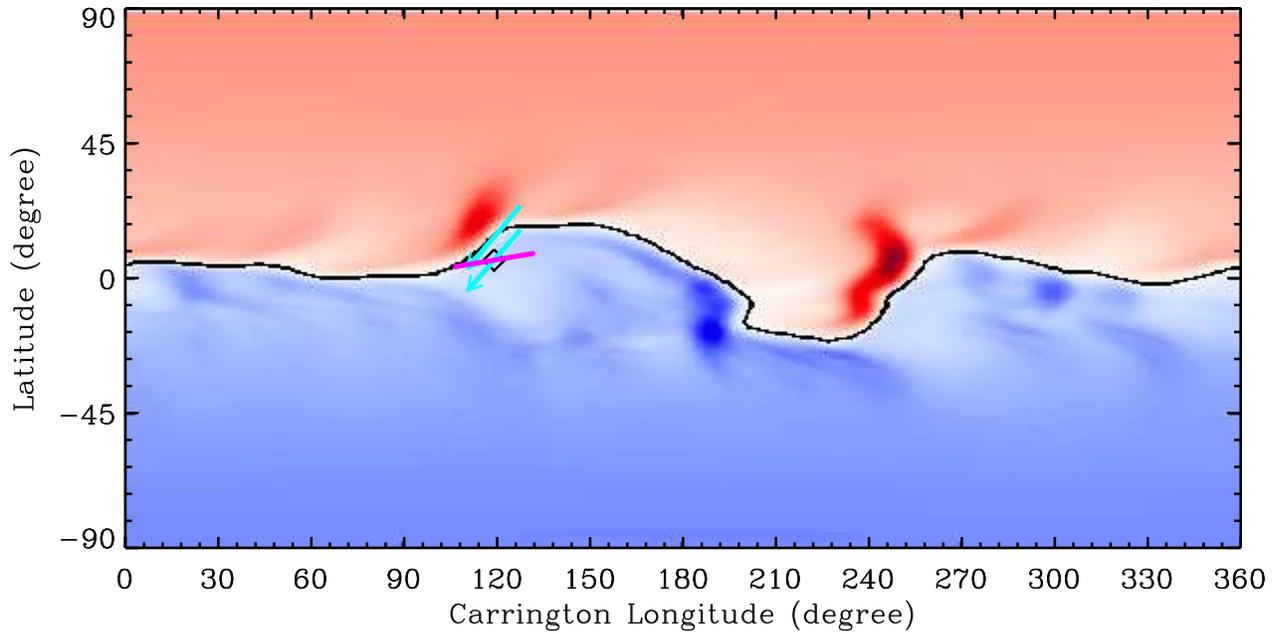}
\caption{\label{f10}The heliospheric current sheet map at 1 AU at 14:10 UT on 2018 August 25 (adapted from \url{http://www.predsci.com/mhdweb/summary_plots.php}). Red (blue) represents the positive (negative) radial component of the heliospheric magnetic field. The tilt angle of the CME flux rope derived from the GCS model and the orientation of the flux rope at 1 AU are indicated by the magenta solid line and the cyan arrow, respectively. The upper cyan arrow is a translation from the lower cyan arrow to the heliospheric current sheet along the same longitude.}
\end{figure}


\clearpage
\bibliographystyle{aasjournal}
\bibliography{article}

\begin{thebibliography}{}
\expandafter\ifx\csname natexlab\endcsname\relax\def\natexlab#1{#1}\fi

\bibitem[{{Aly}(1990)}]{1990Aly}
{Aly}, J.~J. 1990, Computer Physics Communications, 59, 13

\bibitem[{{Antiochos} {et~al.}(1999){Antiochos}, {DeVore}, \&
  {Klimchuk}}]{1999Antiochos}
{Antiochos}, S.~K., {DeVore}, C.~R., \& {Klimchuk}, J.~A. 1999, \apj, 510, 485

\bibitem[{{Burlaga} {et~al.}(1981){Burlaga}, {Sittler}, {Mariani}, \&
  {Schwenn}}]{1981Burlaga}
{Burlaga}, L., {Sittler}, E., {Mariani}, F., \& {Schwenn}, R. 1981, \jgr, 86,
  6673

\bibitem[{{Davies} {et~al.}(2009){Davies}, {Harrison}, {Rouillard}, {Sheeley},
  {Perry}, {Bewsher}, {Davis}, {Eyles}, {Crothers}, \& {Brown}}]{2009Davies}
{Davies}, J.~A., {Harrison}, R.~A., {Rouillard}, A.~P., {et~al.} 2009, \grl,
  36, L02102

\bibitem[{{de Toma} {et~al.}(2005){de Toma}, {Holzer}, {Burkepile}, \&
  {Gilbert}}]{2005deToma}
{de Toma}, G., {Holzer}, T.~E., {Burkepile}, J.~T., \& {Gilbert}, H.~R. 2005,
  \apj, 621, 1109

\bibitem[{{D{\'e}moulin} \& {Aulanier}(2010)}]{2010Demoulin}
{D{\'e}moulin}, P., \& {Aulanier}, G. 2010, \apj, 718, 1388

\bibitem[{{Domingo} {et~al.}(1995){Domingo}, {Fleck}, \&
  {Poland}}]{1995Domingo}
{Domingo}, V., {Fleck}, B., \& {Poland}, A.~I. 1995, \solphys, 162, 1

\bibitem[{{Echer} {et~al.}(2008){Echer}, {Gonzalez}, {Tsurutani}, \&
  {Gonzalez}}]{2008Echer}
{Echer}, E., {Gonzalez}, W.~D., {Tsurutani}, B.~T., \& {Gonzalez}, A.~L.~C.
  2008, Journal of Geophysical Research (Space Physics), 113, A05221

\bibitem[{{Echer} {et~al.}(2009){Echer}, {Tsurutani}, \&
  {Guarnieri}}]{2009Echer}
{Echer}, E., {Tsurutani}, B.~T., \& {Guarnieri}, F.~L. 2009, Advances in Space
  Research, 44, 615

\bibitem[{{Feynman} \& {Martin}(1995)}]{1995Feynman}
{Feynman}, J., \& {Martin}, S.~F. 1995, \jgr, 100, 3355

\bibitem[{{Forbes} \& {Isenberg}(1991)}]{1991Forbes}
{Forbes}, T.~G., \& {Isenberg}, P.~A. 1991, \apj, 373, 294

\bibitem[{Gilbert {et~al.}(2000)Gilbert, Holzer, Burkepile, \&
  Hundhausen}]{2000Gilbert}
Gilbert, H.~R., Holzer, T.~E., Burkepile, J.~T., \& Hundhausen, A.~J. 2000, The
  Astrophysical Journal, 537, 503

\bibitem[{{Gonzalez} {et~al.}(1994){Gonzalez}, {Joselyn}, {Kamide}, {Kroehl},
  {Rostoker}, {Tsurutani}, \& {Vasyliunas}}]{1994Gonzalez}
{Gonzalez}, W.~D., {Joselyn}, J.~A., {Kamide}, Y., {et~al.} 1994, \jgr, 99,
  5771

\bibitem[{{Gopalswamy} {et~al.}(2000){Gopalswamy}, {Lara}, {Lepping}, {Kaiser},
  {Berdichevsky}, \& {St. Cyr}}]{2000Gopalswamy}
{Gopalswamy}, N., {Lara}, A., {Lepping}, R.~P., {et~al.} 2000, \grl, 27, 145

\bibitem[{{Gopalswamy} {et~al.}(2003){Gopalswamy}, {Shimojo}, {Lu}, {Yashiro},
  {Shibasaki}, \& {Howard}}]{2003Gopalswamy}
{Gopalswamy}, N., {Shimojo}, M., {Lu}, W., {et~al.} 2003, \apj, 586, 562

\bibitem[{{Gopalswamy} {et~al.}(2001){Gopalswamy}, {Yashiro}, {Kaiser},
  {Howard}, \& {Bougeret}}]{2001bGopalswamy}
{Gopalswamy}, N., {Yashiro}, S., {Kaiser}, M.~L., {Howard}, R.~A., \&
  {Bougeret}, J.-L. 2001, \apjl, 548, L91

\bibitem[{{Green} {et~al.}(2007){Green}, {Kliem}, {T{\"o}r{\"o}k}, {van
  Driel-Gesztelyi}, \& {Attrill}}]{2007Green}
{Green}, L.~M., {Kliem}, B., {T{\"o}r{\"o}k}, T., {van Driel-Gesztelyi}, L., \&
  {Attrill}, G.~D.~R. 2007, \solphys, 246, 365

\bibitem[{{Hau} \& {Sonnerup}(1999)}]{1999Hau}
{Hau}, L.-N., \& {Sonnerup}, B.~U.~{\"O}. 1999, \jgr, 104, 6899

\bibitem[{{He} {et~al.}(2018){He}, {Liu}, {Hu}, {Wang}, \& {Zhao}}]{2018He}
{He}, W., {Liu}, Y.~D., {Hu}, H., {Wang}, R., \& {Zhao}, X. 2018, \apj, 860, 78

\bibitem[{{Hori} \& {Culhane}(2002)}]{2002Hori}
{Hori}, K., \& {Culhane}, J.~L. 2002, in Multi-Wavelength Observations of
  Coronal Structure and Dynamics, ed. P.~C.~H. {Martens} \& D.~{Cauffman},
  Vol.~10, 305

\bibitem[{{Howard} {et~al.}(2008){Howard}, {Moses}, {Vourlidas}, {Newmark},
  {Socker}, {Plunkett}, {Korendyke}, {Cook}, {Hurley}, {Davila}, {Thompson},
  {St Cyr}, {Mentzell}, {Mehalick}, {Lemen}, {Wuelser}, {Duncan}, {Tarbell},
  {Wolfson}, {Moore}, {Harrison}, {Waltham}, {Lang}, {Davis}, {Eyles},
  {Mapson-Menard}, {Simnett}, {Halain}, {Defise}, {Mazy}, {Rochus}, {Mercier},
  {Ravet}, {Delmotte}, {Auchere}, {Delaboudiniere}, {Bothmer}, {Deutsch},
  {Wang}, {Rich}, {Cooper}, {Stephens}, {Maahs}, {Baugh}, {McMullin}, \&
  {Carter}}]{2008Howard}
{Howard}, R.~A., {Moses}, J.~D., {Vourlidas}, A., {et~al.} 2008, \ssr, 136, 67

\bibitem[{{Hu} {et~al.}(2016){Hu}, {Liu}, {Wang}, {M{\"o}stl}, \&
  {Yang}}]{2016Hu}
{Hu}, H., {Liu}, Y.~D., {Wang}, R., {M{\"o}stl}, C., \& {Yang}, Z. 2016, \apj,
  829, 97

\bibitem[{{Hu} {et~al.}(2017){Hu}, {Liu}, {Wang}, {Zhao}, {Zhu}, \&
  {Yang}}]{2017Hu}
{Hu}, H., {Liu}, Y.~D., {Wang}, R., {et~al.} 2017, \apj, 840, 76

\bibitem[{{Hu} \& {Sonnerup}(2002)}]{2002Hu}
{Hu}, Q., \& {Sonnerup}, B.~U.~{\"O}. 2002, Journal of Geophysical Research
  (Space Physics), 107, 1142

\bibitem[{{Kaiser} {et~al.}(2008){Kaiser}, {Kucera}, {Davila}, {St.~Cyr},
  {Guhathakurta}, \& {Christian}}]{2008Kaiser}
{Kaiser}, M.~L., {Kucera}, T.~A., {Davila}, J.~M., {et~al.} 2008, \ssr, 136, 5

\bibitem[{{Kataoka} {et~al.}(2015){Kataoka}, {Shiota}, {Kilpua}, \&
  {Keika}}]{2015Kataoka}
{Kataoka}, R., {Shiota}, D., {Kilpua}, E., \& {Keika}, K. 2015, \grl, 42, 5155

\bibitem[{{Kay} {et~al.}(2015){Kay}, {Opher}, \& {Evans}}]{2015Kay}
{Kay}, C., {Opher}, M., \& {Evans}, R.~M. 2015, \apj, 805, 168

\bibitem[{{Kilpua} {et~al.}(2009){Kilpua}, {Pomoell}, {Vourlidas}, {Vainio},
  {Luhmann}, {Li}, {Schroeder}, {Galvin}, \& {Simunac}}]{2009Kilpua}
{Kilpua}, E.~K.~J., {Pomoell}, J., {Vourlidas}, A., {et~al.} 2009, Annales
  Geophysicae, 27, 4491

\bibitem[{{Kliem} \& {T{\"o}r{\"o}k}(2006)}]{2006Kliem}
{Kliem}, B., \& {T{\"o}r{\"o}k}, T. 2006, Physical Review Letters, 96, 255002

\bibitem[{{Lemen} {et~al.}(2012){Lemen}, {Title}, {Akin}, {Boerner}, {Chou},
  {Drake}, {Duncan}, {Edwards}, {Friedlaender}, {Heyman}, {Hurlburt}, {Katz},
  {Kushner}, {Levay}, {Lindgren}, {Mathur}, {McFeaters}, {Mitchell}, {Rehse},
  {Schrijver}, {Springer}, {Stern}, {Tarbell}, {Wuelser}, {Wolfson}, {Yanari},
  {Bookbinder}, {Cheimets}, {Caldwell}, {Deluca}, {Gates}, {Golub}, {Park},
  {Podgorski}, {Bush}, {Scherrer}, {Gummin}, {Smith}, {Auker}, {Jerram},
  {Pool}, {Soufli}, {Windt}, {Beardsley}, {Clapp}, {Lang}, \&
  {Waltham}}]{2012Lemen}
{Lemen}, J.~R., {Title}, A.~M., {Akin}, D.~J., {et~al.} 2012, \solphys, 275, 17

\bibitem[{{Liu} {et~al.}(2010{\natexlab{a}}){Liu}, {Davies}, {Luhmann},
  {Vourlidas}, {Bale}, \& {Lin}}]{2010aLiu}
{Liu}, Y., {Davies}, J.~A., {Luhmann}, J.~G., {et~al.} 2010{\natexlab{a}},
  \apjl, 710, L82

\bibitem[{{Liu} {et~al.}(2010{\natexlab{b}}){Liu}, {Thernisien}, {Luhmann},
  {Vourlidas}, {Davies}, {Lin}, \& {Bale}}]{2010bLiu}
{Liu}, Y., {Thernisien}, A., {Luhmann}, J.~G., {et~al.} 2010{\natexlab{b}},
  \apj, 722, 1762

\bibitem[{{Liu} {et~al.}(2008){Liu}, {Luhmann}, {M{\"u}ller-Mellin},
  {Schroeder}, {Wang}, {Lin}, {Bale}, {Li}, {Acu{\~n}a}, \&
  {Sauvaud}}]{2008Liu}
{Liu}, Y., {Luhmann}, J.~G., {M{\"u}ller-Mellin}, R., {et~al.} 2008, \apj, 689,
  563

\bibitem[{{Liu} {et~al.}(2018){Liu}, {Liu}, {Hu}, {Wang}, \&
  {Zhao}}]{2018LiuYi}
{Liu}, Y.~A., {Liu}, Y.~D., {Hu}, H., {Wang}, R., \& {Zhao}, X. 2018, \apj,
  854, 126

\bibitem[{{Liu} {et~al.}(2016){Liu}, {Hu}, {Wang}, {Luhmann}, {Richardson},
  {Yang}, \& {Wang}}]{2016Liu}
{Liu}, Y.~D., {Hu}, H., {Wang}, C., {et~al.} 2016, \apjs, 222, 23

\bibitem[{{Liu} {et~al.}(2015){Liu}, {Hu}, {Wang}, {Yang}, {Zhu}, {Liu},
  {Luhmann}, \& {Richardson}}]{2015Liu}
{Liu}, Y.~D., {Hu}, H., {Wang}, R., {et~al.} 2015, \apjl, 809, L34

\bibitem[{{Liu} {et~al.}(2013){Liu}, {Luhmann}, {Lugaz}, {M{\"o}stl}, {Davies},
  {Bale}, \& {Lin}}]{2013Liu}
{Liu}, Y.~D., {Luhmann}, J.~G., {Lugaz}, N., {et~al.} 2013, \apj, 769, 45

\bibitem[{{Liu} {et~al.}(2014){Liu}, {Luhmann}, {Kajdi{\v c}}, {Kilpua},
  {Lugaz}, {Nitta}, {M{\"o}stl}, {Lavraud}, {Bale}, {Farrugia}, \&
  {Galvin}}]{2014Liu}
{Liu}, Y.~D., {Luhmann}, J.~G., {Kajdi{\v c}}, P., {et~al.} 2014, Nature
  Communications, 5, 3481

\bibitem[{{Lopez}(1987)}]{1987Lopez}
{Lopez}, R.~E. 1987, \jgr, 92, 11189

\bibitem[{{Lugaz} {et~al.}(2012){Lugaz}, {Farrugia}, {Davies}, {M{\"o}stl},
  {Davis}, {Roussev}, \& {Temmer}}]{2012Lugaz}
{Lugaz}, N., {Farrugia}, C.~J., {Davies}, J.~A., {et~al.} 2012, \apj, 759, 68

\bibitem[{{Lugaz} {et~al.}(2017){Lugaz}, {Farrugia}, {Winslow}, {Small},
  {Manion}, \& {Savani}}]{2017Lugaz}
{Lugaz}, N., {Farrugia}, C.~J., {Winslow}, R.~M., {et~al.} 2017, \apj, 848, 75

\bibitem[{{Lugaz} {et~al.}(2009){Lugaz}, {Vourlidas}, \& {Roussev}}]{2009Lugaz}
{Lugaz}, N., {Vourlidas}, A., \& {Roussev}, I.~I. 2009, Annales Geophysicae,
  27, 3479

\bibitem[{{Lynch} {et~al.}(2010){Lynch}, {Li}, {Thernisien}, {Robbrecht},
  {Fisher}, {Luhmann}, \& {Vourlidas}}]{2010Lynch}
{Lynch}, B.~J., {Li}, Y., {Thernisien}, A.~F.~R., {et~al.} 2010, Journal of
  Geophysical Research (Space Physics), 115, A07106

\bibitem[{{MacQueen} {et~al.}(1986){MacQueen}, {Hundhausen}, \&
  {Conover}}]{1986MacQueen}
{MacQueen}, R.~M., {Hundhausen}, A.~J., \& {Conover}, C.~W. 1986, \jgr, 91, 31

\bibitem[{{Manchester} {et~al.}(2017){Manchester}, {Kilpua}, {Liu}, {Lugaz},
  {Riley}, {T{\"o}r{\"o}k}, \& {Vr{\v{s}}nak}}]{2017Manchester}
{Manchester}, W., {Kilpua}, E. K.~J., {Liu}, Y.~D., {et~al.} 2017, \ssr, 212,
  1159

\bibitem[{{Marubashi} {et~al.}(2015){Marubashi}, {Akiyama}, {Yashiro},
  {Gopalswamy}, {Cho}, \& {Park}}]{2015Marubashi}
{Marubashi}, K., {Akiyama}, S., {Yashiro}, S., {et~al.} 2015, \solphys, 290,
  1371

\bibitem[{Meng {et~al.}(2019)Meng, Tsurutani, \& Mannucci}]{2019MengXing}
Meng, X., Tsurutani, B.~T., \& Mannucci, A.~J. 2019, Journal of Geophysical
  Research: Space Physics, 124, 3926

\bibitem[{{M{\"o}stl} {et~al.}(2009){M{\"o}stl}, {Farrugia}, {Miklenic},
  {Temmer}, {Galvin}, {Luhmann}, {Kilpua}, {Leitner}, {Nieves-Chinchilla},
  {Veronig}, \& {Biernat}}]{2009Mostl}
{M{\"o}stl}, C., {Farrugia}, C.~J., {Miklenic}, C., {et~al.} 2009, Journal of
  Geophysical Research (Space Physics), 114, A04102

\bibitem[{Munro {et~al.}(1979)Munro, Gosling, Hildner, MacQueen, Poland, \&
  Ross}]{1979Munro}
Munro, R.~H., Gosling, J.~T., Hildner, E., {et~al.} 1979, Solar Physics, 61,
  201

\bibitem[{{Priest} {et~al.}(1989){Priest}, {Hood}, \& {Anzer}}]{1989Priest}
{Priest}, E.~R., {Hood}, A.~W., \& {Anzer}, U. 1989, \apj, 344, 1010

\bibitem[{{Rouillard} {et~al.}(2010){Rouillard}, {Lavraud}, {Sheeley},
  {Davies}, {Burlaga}, {Savani}, {Jacquey}, \& {Forsyth}}]{2010Rouillard}
{Rouillard}, A.~P., {Lavraud}, B., {Sheeley}, N.~R., {et~al.} 2010, \apj, 719,
  1385

\bibitem[{{Rust} {et~al.}(2005){Rust}, {Anderson}, {Andrews}, {Acu{\~n}a},
  {Russell}, {Schuck}, \& {Mulligan}}]{2005Rust}
{Rust}, D.~M., {Anderson}, B.~J., {Andrews}, M.~D., {et~al.} 2005, \apj, 621,
  524

\bibitem[{{Schou} {et~al.}(2012){Schou}, {Scherrer}, {Bush}, {Wachter},
  {Couvidat}, {Rabello-Soares}, {Bogart}, {Hoeksema}, {Liu}, {Duvall}, {Akin},
  {Allard}, {Miles}, {Rairden}, {Shine}, {Tarbell}, {Title}, {Wolfson},
  {Elmore}, {Norton}, \& {Tomczyk}}]{2012Schou}
{Schou}, J., {Scherrer}, P.~H., {Bush}, R.~I., {et~al.} 2012, \solphys, 275,
  229

\bibitem[{{Sheeley} {et~al.}(2008){Sheeley}, {Herbst}, {Palatchi}, {Wang},
  {Howard}, {Moses}, {Vourlidas}, {Newmark}, {Socker}, {Plunkett}, {Korendyke},
  {Burlaga}, {Davila}, {Thompson}, {St.~Cyr}, {Harrison}, {Davis}, {Eyles},
  {Halain}, {Wang}, {Rich}, {Battams}, {Esfandiari}, \&
  {Stenborg}}]{2008Sheeley}
{Sheeley}, Jr., N.~R., {Herbst}, A.~D., {Palatchi}, C.~A., {et~al.} 2008, \apj,
  675, 853

\bibitem[{{Sterling} {et~al.}(2007){Sterling}, {Harra}, \&
  {Moore}}]{2007Sterling}
{Sterling}, A.~C., {Harra}, L.~K., \& {Moore}, R.~L. 2007, \apj, 669, 1359

\bibitem[{{Sterling} \& {Hudson}(1997)}]{1997Sterling}
{Sterling}, A.~C., \& {Hudson}, H.~S. 1997, \apjl, 491, L55

\bibitem[{{Tang} {et~al.}(1989){Tang}, {Tsurutani}, {Gonzalez}, {Akasofu}, \&
  {Smith}}]{1989Tang}
{Tang}, F., {Tsurutani}, B.~T., {Gonzalez}, W.~D., {Akasofu}, S.~I., \&
  {Smith}, E.~J. 1989, \jgr, 94, 3535

\bibitem[{{Thernisien} {et~al.}(2009){Thernisien}, {Vourlidas}, \&
  {Howard}}]{2009Thernisien}
{Thernisien}, A., {Vourlidas}, A., \& {Howard}, R.~A. 2009, \solphys, 256, 111

\bibitem[{{Thernisien} {et~al.}(2006){Thernisien}, {Howard}, \&
  {Vourlidas}}]{2006Thernisien}
{Thernisien}, A.~F.~R., {Howard}, R.~A., \& {Vourlidas}, A. 2006, \apj, 652,
  763

\bibitem[{{T{\"o}r{\"o}k} {et~al.}(2004){T{\"o}r{\"o}k}, {Kliem}, \&
  {Titov}}]{2004Torok}
{T{\"o}r{\"o}k}, T., {Kliem}, B., \& {Titov}, V.~S. 2004, \aap, 413, L27

\bibitem[{{Tsurutani} {et~al.}(1988){Tsurutani}, {Gonzalez}, {Tang}, {Akasofu},
  \& {Smith}}]{1988Tsurutani}
{Tsurutani}, B.~T., {Gonzalez}, W.~D., {Tang}, F., {Akasofu}, S.~I., \&
  {Smith}, E.~J. 1988, \jgr, 93, 8519

\bibitem[{{Tsurutani} {et~al.}(2004){Tsurutani}, {Gonzalez}, {Zhou}, {Lepping},
  \& {Bothmer}}]{2004Tsurutani}
{Tsurutani}, B.~T., {Gonzalez}, W.~D., {Zhou}, X.-Y., {Lepping}, R.~P., \&
  {Bothmer}, V. 2004, Journal of Atmospheric and Solar-Terrestrial Physics, 66,
  147

\bibitem[{{Tsurutani} {et~al.}(2011){Tsurutani}, {Lakhina}, {Verkhoglyadova},
  {Gonzalez}, {Echer}, \& {Guarnieri}}]{2011Tsurutani}
{Tsurutani}, B.~T., {Lakhina}, G.~S., {Verkhoglyadova}, O.~P., {et~al.} 2011,
  Journal of Atmospheric and Solar-Terrestrial Physics, 73, 5

\bibitem[{{Vemareddy} {et~al.}(2016){Vemareddy}, {M{\"o}stl}, {Amerstorfer},
  {Mishra}, {Farrugia}, \& {Leitner}}]{2016Vemareddy}
{Vemareddy}, P., {M{\"o}stl}, C., {Amerstorfer}, T., {et~al.} 2016, \apj, 828,
  12

\bibitem[{{Vourlidas} {et~al.}(2011){Vourlidas}, {Colaninno},
  {Nieves-Chinchilla}, \& {Stenborg}}]{2011Vourlidas}
{Vourlidas}, A., {Colaninno}, R., {Nieves-Chinchilla}, T., \& {Stenborg}, G.
  2011, \apj, 733, L23

\bibitem[{{Wang} {et~al.}(2015){Wang}, {Liu}, {Dai}, {Yang}, {Huang}, \&
  {Hu}}]{2015Wang}
{Wang}, R., {Liu}, Y.~D., {Dai}, X., {et~al.} 2015, \apj, 814, 80

\bibitem[{{Xie} {et~al.}(2009){Xie}, {St.~Cyr}, {Gopalswamy}, {Yashiro},
  {Krall}, {Kramar}, \& {Davila}}]{2009Xie}
{Xie}, H., {St.~Cyr}, O.~C., {Gopalswamy}, N., {et~al.} 2009, \solphys, 259,
  143

\bibitem[{{Yurchyshyn}(2008)}]{2008Yurchyshyn}
{Yurchyshyn}, V. 2008, \apjl, 675, L49

\bibitem[{{Yurchyshyn} {et~al.}(2007){Yurchyshyn}, {Hu}, {Lepping}, {Lynch}, \&
  {Krall}}]{2007Yurchyshyn}
{Yurchyshyn}, V., {Hu}, Q., {Lepping}, R.~P., {Lynch}, B.~J., \& {Krall}, J.
  2007, Advances in Space Research, 40, 1821

\bibitem[{{Zhao} {et~al.}(2017){Zhao}, {Liu}, {Hu}, \& {Wang}}]{2017Zhao}
{Zhao}, X., {Liu}, Y.~D., {Hu}, H., \& {Wang}, R. 2017, \apj, 837, 4

\end{thebibliography}

\end{document}